\documentclass{article}%
\usepackage{geometry}
\usepackage[doublespacing]{setspace}
\usepackage{amsmath}
\usepackage{amsfonts}
\usepackage{amssymb}
\usepackage{graphicx}%
\begin{document}

\begin{center}
{\large Dyadic Green's Functions and Guided Surface Waves for a Surface
Conductivity Model of Graphene }

\bigskip

George W. Hanson\linebreak

Department of Electrical Engineering, University of
Wisconsin-Milwaukee\linebreak

3200 N. Cramer St., Milwaukee, Wisconsin 53211, USA.
\end{center}

\section{Abstract}

An exact solution is obtained for the electromagnetic field due to an electric
current in the presence of a surface conductivity model of graphene. The
graphene is represented by an infinitesimally-thin, local and isotropic
two-sided conductivity surface. The field is obtained in terms of dyadic
Green's functions represented as Sommerfeld integrals. The solution of
plane-wave reflection and transmission is presented, and surface wave
propagation along graphene is studied via the poles of the Sommerfeld
integrals. For isolated graphene characterized by complex surface conductivity
$\sigma=\sigma^{\prime}+j\sigma^{\prime\prime}$, a proper transverse-electric
(TE) surface wave exists if and only if $\sigma^{\prime\prime}>0$ (associated
with interband conductivity), and a proper transverse-magnetic (TM) surface
wave exists for $\sigma^{\prime\prime}<0$ (associated with intraband
conductivity). By tuning the chemical potential at infrared frequencies, the
sign of $\sigma^{\prime\prime}$ can be varied, allowing for some control over
surface wave properties.\newpage

\section{Introduction}

Fundamental properties and potential applications of carbon-based structures
are of interest in emerging nanoelectronic applications. Particularly
promising is graphene, which is a planar atomic layer of carbon atoms bonded
in a hexagonal structure. Graphene is the two-dimensional version of graphite,
and a single-wall carbon nanotube can be thought of as graphene rolled into a
tube \cite{SDD03}.

Only recently has it become possible to fabricate ultrathin graphite,
consisting of only a few graphene layers \cite{BS2004}--\cite{ZSPK2005}, and
actual graphene \cite{BSL2006}--\cite{HF2006}. In graphene, the
energy-momentum relationship for electrons is linear over a wide range of
energies, rather then quadratic, so that electrons in graphene behave as
massless relativistic particles (Dirac fermions) with an energy-independent
velocity. Graphene's band structure, together with it's extreme thinness,
leads to a pronounced electric field effect \cite{NGM2004}, \cite{OYFG2006},
which is the variation of a material's carrier concentration with
electrostatic gating. This is the governing principle behind traditional
semiconductor device operation, and therefore this effect in graphene is
particularly promising for the development of ultrathin carbon nanoelectronic
devices. Although the electric field effect also occurs in atomically thin
metal films, these tend to be thermodynamically unstable, and do not form
continuous layers with good transport properties. In contrast, graphene is
stable, and, like its cylindrical carbon nanotube versions, can exhibit
ballistic transport over at least submicron distances \cite{NGM2004}.
Furthermore, it has been shown that graphene's conductance has a minimum,
non-zero value associated with the conductance quantum, even when charge
carrier concentrations vanish \cite{NGM2005}.

In this work, the interaction of an electromagnetic current source and
graphene is considered. The electromagnetic fields are governed by Maxwell's
equations, and the graphene is represented by a conductivity surface
\cite{SMLYG99} that must arise from a microscopic quantum-dynamical model, or
from measurement. The method assumes laterally infinite graphene residing at
the interface between two dielectrics, in which case classical Maxwell's
equations are solved exactly for an arbitrary electrical current. Related
phenomena are discussed, such as plane-wave reflection and transmission
through graphene \cite{FPPRB2007}, and surface wave excitation and guidance,
which is relevant to high-frequency electronic applications. It is found that
the relative importance of the interband and intraband contributions to the
conductivity dictate surface wave behavior \cite{MZPRL2007}, and that surface
wave propagation can be controlled by varying the chemical potential.

Although at this time only graphene samples with lateral dimensions on the
order of tens of microns have been fabricated, the infinite sheet model
provides a first step in analyzing electromagnetic properties of graphene. It
is also relevant to sufficiently large finite-sized sheets, assuming that
electronic edge effects and electromagnetic edge diffraction can be ignored.
In the following all units are in the SI system, and the time variation
(suppressed) is $e^{j\omega t}$, where $j$ is the imaginary unit.

\section{Formulation of the Model}

\subsection{Electronic Model of Graphene}

Figure 1 depicts laterally-infinite graphene lying in the $x-z$ plane at the
interface between two different mediums characterized by $\mu_{1}%
,\varepsilon_{1}$ for $y\geq0$ and $\mu_{2},\varepsilon_{2}$ for $y<0$, where
all material parameters may be complex-valued.

The graphene is modeled as an infinitesimally-thin, local two-sided surface
characterized by a surface conductivity $\sigma\left(  \omega,\mu_{c}%
,\Gamma,T\right)  $, where $\omega$ is radian frequency, $\mu_{c}$ is chemical
potential, $\Gamma$ is a phenomenological scattering rate that is assumed to
be independent of energy $\varepsilon$, and $T$ is temperature. The
conductivity of graphene has been considered in several recent works
\cite{FPPRB2007}--\cite{FV2006}, and here we use the expression resulting from
the Kubo formula \cite{GSC2007},%
\begin{align}
&  \sigma\left(  \omega,\mu_{c},\Gamma,T\right)  =\frac{je^{2}\left(
\omega-j2\Gamma\right)  }{\pi\hslash^{2}}\left[  \frac{1}{\left(
\omega-j2\Gamma\right)  ^{2}}\int_{0}^{\infty}\varepsilon\left(
\frac{\partial f_{d}\left(  \varepsilon\right)  }{\partial\varepsilon}%
-\frac{\partial f_{d}\left(  -\varepsilon\right)  }{\partial\varepsilon
}\right)  d\varepsilon\right. \label{GSC}\\
&  \ \ \ \ \ \ \ \ \ \ \ \ \ \ \ \ -\left.  \int_{0}^{\infty}\frac
{f_{d}\left(  -\varepsilon\right)  -f_{d}\left(  \varepsilon\right)  }{\left(
\omega-j2\Gamma\right)  ^{2}-4\left(  \varepsilon/\hslash\right)  ^{2}%
}d\varepsilon\right] \nonumber
\end{align}
where $e$ is the charge of an electron, $\hslash=h/2\pi$ is the reduced
Planck's constant, $f_{d}\left(  \varepsilon\right)  =\left(  e^{\left(
\varepsilon-\mu_{c}\right)  /k_{B}T}+1\right)  ^{-1}$ is the Fermi-Dirac
distribution, and $k_{B}$ is Boltzmann's constant. We assume that no external
magnetic field is present, and so the local conductivity is isotropic (i.e.,
there is no Hall conductivity). The first term in (\ref{GSC}) is due to
intraband contributions, and the second term to interband contributions.

For an isolated graphene sheet the chemical potential $\mu_{c}$ is determined
by the carrier density $n_{s}$,%
\begin{equation}
n_{s}=\frac{2}{\pi\hslash^{2}v_{F}^{2}}\int_{0}^{\infty}\varepsilon\left(
f_{d}\left(  \varepsilon\right)  -f_{d}\left(  \varepsilon+2\mu_{c}\right)
\right)  d\varepsilon, \label{cd}%
\end{equation}
where $v_{F}\simeq9.5\times10^{5}$ m/s is the Fermi velocity. The carrier
density can be controlled by application of a gate voltage and/or chemical
doping. For the undoped, ungated case at $T=0$ K, $n_{s}=\mu_{c}=0$.

The intraband term in (\ref{GSC}) can be evaluated as%
\begin{equation}
\sigma_{intra}\left(  \omega,\mu_{c},\Gamma,T\right)  =-j\frac{e^{2}k_{B}%
T}{\pi\hslash^{2}\left(  \omega-j2\Gamma\right)  }\left(  \frac{\mu_{c}}%
{k_{B}T}+2\ln\left(  e^{-\frac{\mu_{c}}{k_{B}T}}+1\right)  \right)  .
\label{intraLR}%
\end{equation}
For the case $\mu_{c}=0$, (\ref{intraLR}) was first derived in
\cite{wallace1947} for graphite (with the addition of a factor to account for
the interlayer separation between graphene planes), and corresponds to the
intraband conductivity of a single-wall carbon nanotube in the limit of
infinite radius \cite{SMLYG99}. With $\sigma=\sigma^{\prime}+j\sigma
^{\prime\prime}$, it can be seen that $\sigma_{intra}^{\prime}\geq0$ and
$\sigma_{intra}^{\prime\prime}<0$. As will be discussed later, the imaginary
part of conductivity plays an important role in the propagation of surface
waves guided by the graphene sheet \cite{MZPRL2007}.

The interband conductivity can be approximated for $k_{B}T\ll\left\vert
\mu_{c}\right\vert ,\hslash\omega$ as \cite{GSC2007sum},%
\begin{equation}
\sigma_{inter}\left(  \omega,\mu_{c},\Gamma,0\right)  \simeq\frac{-je^{2}%
}{4\pi\hslash}\ln\left(  \frac{2\left\vert \mu_{c}\right\vert -\left(
\omega-j2\Gamma\right)  \hslash}{2\left\vert \mu_{c}\right\vert +\left(
\omega-j2\Gamma\right)  \hslash}\right)  , \label{inter}%
\end{equation}
such that for $\Gamma=0$ and $2\left\vert \mu_{c}\right\vert >\hslash\omega$,
$\sigma_{inter}=j\sigma_{inter}^{\prime\prime}$ with $\sigma_{inter}%
^{\prime\prime}>0$. For $\Gamma=0$ and $2\left\vert \mu_{c}\right\vert
<\hslash\omega$, $\sigma_{inter}$ is complex-valued, with \cite{sigma}%
\begin{equation}
\sigma_{inter}^{\prime}=\frac{\pi e^{2}}{2h}=\sigma_{\min}=6.085\times
10^{-5}\text{ (S),} \label{smin}%
\end{equation}
and $\sigma_{inter}^{\prime\prime}>0$ for $\mu_{c}\neq0$.

\subsection{Dyadic Green's Function for a Surface Model of Graphene}

For any planarly layered, piecewise-constant medium, the electric and magnetic
fields in region $n$ due to an electric current can be obtained as
\cite{Chew1990}--\cite{Ishimaru}%
\begin{align}
\mathbf{E}^{\left(  n\right)  }\left(  \mathbf{r}\right)   &  =\left(
k_{n}^{2}+\mathbf{\nabla\nabla\cdot}\right)  \mathbf{\pi}^{\left(  n\right)
}\left(  \mathbf{r}\right)  ,\label{ffp}\\
\mathbf{H}^{\left(  n\right)  }\left(  \mathbf{r}\right)   &  =j\omega
\varepsilon_{n}\mathbf{\nabla}\times\mathbf{\pi}^{\left(  n\right)  }\left(
\mathbf{r}\right)  , \label{fpp1}%
\end{align}
where $k_{n}=\omega\sqrt{\mu_{n}\varepsilon_{n}}$ and $\mathbf{\pi}^{\left(
n\right)  }\left(  \mathbf{r}\right)  $ are the wavenumber and electric
Hertzian potential in region $n$, respectively. Assuming that the current
source is in region 1, then%
\begin{align}
\mathbf{\pi}^{\left(  1\right)  }\left(  \mathbf{r}\right)   &  =\mathbf{\pi
}_{1}^{p}\left(  \mathbf{r}\right)  +\mathbf{\pi}_{1}^{s}\left(
\mathbf{r}\right)  =\int_{\Omega}\left\{  \underline{\mathbf{g}}_{1}%
^{p}\left(  \mathbf{r,r}^{\prime}\right)  +\underline{\mathbf{g}}_{1}%
^{s}\left(  \mathbf{r,r}^{\prime}\right)  \right\}  \cdot\frac{\mathbf{J}%
^{\left(  1\right)  }\left(  \mathbf{r}^{\prime}\right)  }{j\omega
\varepsilon_{1}}\,d\Omega^{\prime},\\
\mathbf{\pi}^{\left(  2\right)  }\left(  \mathbf{r}\right)   &  =\mathbf{\pi
}_{2}^{s}\left(  \mathbf{r}\right)  =\int_{\Omega}\underline{\mathbf{g}}%
_{2}^{s}\left(  \mathbf{r,r}^{\prime}\right)  \cdot\frac{\mathbf{J}^{\left(
1\right)  }\left(  \mathbf{r}^{\prime}\right)  }{j\omega\varepsilon_{1}%
}\,d\Omega^{\prime}, \label{hpt}%
\end{align}
where the underscore indicates a dyadic quantity, and where $\Omega$ is the
support of the current. With $y$ parallel to the interface normal, the
principle Greens dyadic can be written as \cite{Chew1990}%
\begin{equation}
\underline{\mathbf{g}}_{1}^{p}\left(  \mathbf{r,r}^{\prime}\right)
=\underline{\mathbf{I}}\,\frac{e^{-jk_{1}R}}{4\pi R}=\underline{\mathbf{I}%
}\,\frac{1}{2\pi}\int_{-\infty}^{\infty}e^{-p_{1}\left\vert y-y^{\prime
}\right\vert }\frac{H_{0}^{\left(  2\right)  }\left(  k_{\rho}\rho\right)
}{4p_{1}}k_{\rho}dk_{\rho} \label{gp4}%
\end{equation}
where%
\begin{align}
p_{n}^{2}  &  =k_{\rho}^{2}-k_{n}^{2},\ \ \ \rho=\sqrt{\left(  x-x^{\prime
}\right)  ^{2}+\left(  z-z^{\prime}\right)  ^{2}},\label{ar}\\
R  &  =\left\vert \mathbf{r-r}^{\prime}\right\vert =\sqrt{\left(  y-y^{\prime
}\right)  ^{2}+\rho^{2}},\nonumber
\end{align}
and where $k_{\rho}$ is a radial wavenumber and $\underline{\mathbf{I}}$ is
the unit dyadic.

The scattered Green's dyadics can be obtained by enforcing the usual
electromagnetic boundary conditions
\begin{align}
\mathbf{y}\times\left(  \mathbf{H}_{1}-\mathbf{H}_{2}\right)   &
=\mathbf{J}_{e}^{s},\nonumber\\
\mathbf{y}\times\left(  \mathbf{E}_{1}-\mathbf{E}_{2}\right)   &
=-\mathbf{J}_{m}^{s}, \label{BCME}%
\end{align}
where $\mathbf{J}_{e}^{s}$ (A/m) and $\mathbf{J}_{m}^{s}$ (V/m) are electric
and magnetic surface currents on the boundary. For a local model of graphene
in the absence of a magnetic field and associated Hall effect conductivity,
$\sigma$ is a scalar. Therefore \cite{SMLYG99},
\begin{align}
J_{e,x}^{s}\left(  x,y=0,z\right)   &  =\sigma E_{x}\left(  x,y=0,z\right)
,\\
J_{e,z}^{s}\left(  x,y=0,z\right)   &  =\sigma E_{z}\left(  x,y=0,z\right)
,\\
\mathbf{J}_{m}^{s}\left(  x,y=0,z\right)   &  =\mathbf{0},
\end{align}
and (\ref{BCME}) becomes
\begin{align}
&  E_{1,\alpha}\left(  y=0^{+}\right)  =E_{2,\alpha}\left(  y=0^{-}\right)
,\ \ \ \alpha=x,z,\\
&  H_{2,x}\left(  y=0^{-}\right)  -H_{1,x}\left(  y=0^{+}\right)  =\sigma
E_{z}\left(  y=0\right)  ,\\
&  H_{2,z}\left(  y=0^{-}\right)  -H_{1,z}\left(  y=0^{+}\right)  =-\sigma
E_{x}\left(  y=0\right)  .
\end{align}
Using (\ref{ffp}), the boundary conditions on the Hertzian potential at
$\left(  x,y=0,z\right)  $ are%
\begin{align}
&  \pi_{1,\alpha}=N^{2}M^{2}\pi_{2,\alpha},\label{ccg1}\\
&  \varepsilon_{1}\pi_{1,y}-\varepsilon_{2}\pi_{2,y}=\frac{\sigma}{j\omega
}\nabla\cdot\mathbf{\pi}_{1}\\
&  \varepsilon_{2}\frac{\partial\pi_{2,\alpha}}{\partial y}-\varepsilon
_{1}\frac{\partial\pi_{1,\alpha}}{\partial y}=\frac{\sigma}{j\omega}k_{1}%
^{2}\pi_{1,\alpha}\\
&  \left(  \frac{\partial\pi_{1,y}}{\partial y}-\frac{\partial\pi_{2,y}%
}{\partial y}\right)  =\left(  1-N^{2}M^{2}\right)  \left(  \frac{\partial
\pi_{2,x}}{\partial x}+\frac{\partial\pi_{2,z}}{\partial z}\right)  ,
\label{ccg4}%
\end{align}
$\alpha=x,z$, where $N^{2}=\varepsilon_{2}/\varepsilon_{1}$ and $M^{2}=\mu
_{2}/\mu_{1}$. In the absence of magnetic contrast (e.g., if $M=1$) and
surface conductivity, boundary conditions (\ref{ccg1})--(\ref{ccg4}) are
identical to the Hertzian potential boundary conditions presented \cite{BN87}.

Enforcing (\ref{ccg1})--(\ref{ccg4}) and following the method described in
\cite{BN87}, the scattered Green's dyadic is found to be%
\begin{equation}
\underline{\mathbf{g}}_{1}^{s}\left(  \mathbf{r,r}^{\prime}\right)
=\widehat{\mathbf{y}}\widehat{\mathbf{y}}~g_{n}^{s}\left(  \mathbf{r,r}%
^{\prime}\right)  +\left(  \widehat{\mathbf{y}}\widehat{\mathbf{x}}%
\frac{\partial}{\partial x}+\widehat{\mathbf{y}}\widehat{\mathbf{z}}%
\frac{\partial}{\partial z}\right)  g_{c}^{s}\left(  \mathbf{r,r}^{\prime
}\right)  +\left(  \widehat{\mathbf{x}}\widehat{\mathbf{x}}~\mathbf{+~}%
\widehat{\mathbf{z}}\widehat{\mathbf{z}}\right)  g_{t}^{s}\left(
\mathbf{r,r}^{\prime}\right)  , \label{g1s}%
\end{equation}
where the Sommerfeld integrals are%
\begin{equation}
g_{\beta}^{s}\left(  \mathbf{r,r}^{\prime}\right)  =\frac{1}{2\pi}%
\int_{-\infty}^{\infty}R_{\beta}\frac{H_{0}^{\left(  2\right)  }\left(
k_{\rho}\rho\right)  e^{-p_{1}\left(  y+y^{\prime}\right)  }}{4p_{1}}k_{\rho
}dk_{\rho}, \label{sgf3}%
\end{equation}
$\beta=t,n,c$, with%
\begin{align}
R_{t}  &  =\frac{M^{2}p_{1}-p_{2}-j\sigma\omega\mu_{2}}{M^{2}p_{1}%
+p_{2}+j\sigma\omega\mu_{2}}=\frac{N^{H}\left(  k_{\rho},\omega\right)
}{Z^{H}\left(  k_{\rho},\omega\right)  },\label{rtcn}\\
R_{n}  &  =\frac{N^{2}p_{1}-p_{2}+\frac{\sigma p_{1}p_{2}}{j\omega
\varepsilon_{1}}}{N^{2}p_{1}+p_{2}+\frac{\sigma p_{1}p_{2}}{j\omega
\varepsilon_{1}}}=\frac{N^{E}\left(  k_{\rho},\omega\right)  }{Z^{E}\left(
k_{\rho},\omega\right)  },\label{rtc2}\\
R_{c}  &  =\frac{2p_{1}\left[  \left(  N^{2}M^{2}-1\right)  +\frac{\sigma
p_{2}M^{2}}{j\omega\varepsilon_{1}}\right]  }{Z^{H}Z^{E}}, \label{rtcn3}%
\end{align}
which reduce to the previously known results as $\sigma\rightarrow0$.

The Green's dyadic for region 2, $\underline{\mathbf{g}}_{2}^{s}\left(
\mathbf{r,r}^{\prime}\right)  $, has the same form as for region 1, although
in (\ref{sgf3}) the replacement%
\begin{equation}
R_{\beta}e^{-p_{1}\left(  y+y^{\prime}\right)  }\rightarrow T_{\beta}%
~e^{p_{2}y}e^{-p_{1}y^{\prime}}%
\end{equation}
must be made, where
\begin{align}
T_{t}  &  =\frac{\left(  1+R_{t}\right)  }{N^{2}M^{2}}=\frac{2p_{1}}%
{N^{2}Z^{H}},\label{r2}\\
T_{n}  &  =\frac{p_{1}\left(  1-R_{n}\right)  }{p_{2}}=\frac{2p_{1}}{Z^{E}},\\
T_{c}  &  =\frac{2p_{1}\left[  \left(  N^{2}M^{2}-1\right)  +\frac{\sigma
p_{1}}{j\omega\varepsilon_{1}}\right]  }{N^{2}Z^{H}Z^{E}}. \label{r4}%
\end{align}

As in the case of a simple dielectric interface, the denominators
$Z^{H,E}\left(  k_{\rho},\omega\right)  =0$ implicate pole singularities in
the spectral plane associated with surface wave phenomena. Furthermore, both
waveparameters $p_{n}=\sqrt{k_{\rho}^{2}-k_{n}^{2}}$, $n=1,2$, lead to branch
points at $k_{\rho}=\pm k_{n}$, and thus the $k_{\rho}$-plane is a
four-sheeted Riemann surface. The standard hyperbolic branch cuts
\cite{Ishimaru} that separate the one proper sheet (where $\operatorname{Re}%
\left(  p_{n}\right)  >0$, such that the radiation\ condition as $\left\vert
y\right\vert \rightarrow\infty$ is satisfied) and the three improper sheets
(where $\operatorname{Re}\left(  p_{n}\right)  <0$) are the same as in the
absence of surface conductivity $\sigma$.

In addition to representing the exact field from a given current, several
interesting electromagnetic aspects of graphene can be obtained from the above relations.

\subsection{Plane-Wave Reflection and Transmission Coefficients}

Normal incidence plane-wave reflection and transmission coefficients can be
obtained from the previous formulation by setting $k_{\rho}=0$ in (\ref{rtcn})
and (\ref{r2}). To see this, consider the current
\begin{equation}
\mathbf{J}^{\left(  1\right)  }\left(  \mathbf{r}\right)  =\widehat
{\mathbf{\alpha}}\frac{j4\pi r_{0}}{\omega\mu_{1}}\delta\left(  \mathbf{r-r}%
_{0}\right)  ,
\end{equation}
where $\widehat{\mathbf{\alpha}}=\widehat{\mathbf{x}}$ or $\widehat
{\mathbf{y}}$, $\mathbf{r}_{0}=\widehat{\mathbf{y}}y_{0}$, and where $y_{0}%
\gg0$. This current leads to a unit-amplitude, $\widehat{\mathbf{\alpha}}%
$-polarized, transverse electromagnetic plane wave, normally-incident on the
interface. The far scattered field in region 1, which is the reflected field,
can be obtained by evaluating the spectral integral (\ref{sgf3}) using the
method of steepest descents, which leads to the reflected field $\mathbf{E}%
^{r}=\widehat{\mathbf{\alpha}}\Gamma e^{-jk_{1}y}$, where the reflection
coefficient is $\Gamma=R_{t}\left(  k_{\rho}=0\right)  $. In a similar manner,
the far scattered field in region 2, the transmitted field, is obtained as
$\mathbf{E}^{t}=\widehat{\mathbf{\alpha}}Te^{jk_{2}y}$, where the transmission
coefficient is $T=\left(  1+\Gamma\right)  $. Therefore,%
\begin{align}
\Gamma &  =\frac{\eta_{2}-\eta_{1}-\sigma\eta_{1}\eta_{2}}{\eta_{2}+\eta
_{1}+\sigma\eta_{1}\eta_{2}},\\
T  &  =\left(  1+R_{t}\right)  =\frac{2\eta_{2}}{\left(  \eta_{2}+\eta
_{1}+\sigma\eta_{1}\eta_{2}\right)  }%
\end{align}
where $\eta_{n}=\sqrt{\mu_{n}/\varepsilon_{n}}$ is the wave impedance in
region $n$. The plane-wave reflection and transmission coefficients obviously
reduce to the correct results for $\sigma=0$, and in the limit $\sigma
\rightarrow\infty$, $\Gamma\rightarrow-1$ and $T\rightarrow0$ as expected.

In the special case $\varepsilon_{1}=\varepsilon_{2}=\varepsilon_{0}$ and
$\mu_{1}=\mu_{2}=\mu_{0}$,
\begin{equation}
\Gamma=-\frac{\frac{\sigma\eta_{0}}{2}}{1+\frac{\sigma\eta_{0}}{2}%
},\ \ T=\frac{1}{\left(  1+\frac{\sigma\eta_{0}}{2}\right)  },
\end{equation}
where $\eta_{0}=\sqrt{\mu_{0}/\varepsilon_{0}}\simeq377$ ohms. The reflection
coefficient agrees with the result presented in \cite{FPPRB2007} for normal incidence.

\subsection{Surface Waves Guided by Graphene}

Pole singularities in the Sommerfeld integrals represent discrete surface
waves guided by the medium \cite{Chew1990}--\cite{Ishimaru}. From
(\ref{rtcn})--(\ref{rtcn3}) and (\ref{r2})--(\ref{r4}), the dispersion
equation for surface waves that are transverse-electric (TE) to the
propagation direction $\rho$ (also known as $H$-waves) is%
\begin{equation}
Z^{H}\left(  k_{\rho},\omega\right)  =M^{2}p_{1}+p_{2}+j\sigma\omega\mu_{2}=0,
\label{zh}%
\end{equation}
whereas for transverse-magnetic (TM) waves ($E$-waves),%
\begin{equation}
Z^{E}\left(  k_{\rho},\omega\right)  =N^{2}p_{1}+p_{2}+\frac{\sigma p_{1}%
p_{2}}{j\omega\varepsilon_{1}}=0. \label{ze}%
\end{equation}
In the limit that $\varepsilon_{1}=\varepsilon_{2}=\varepsilon_{0}$, $Z^{H,E}$
agree with TE and TM dispersion equations in \cite{MZPRL2007}

The surface wave field can be obtained from the residue contribution of the
Sommerfeld integrals. For example, the electric field in region $1$ associated
with the surface wave excited by a Hertzian dipole current $\mathbf{J}\left(
\mathbf{r}\right)  =\widehat{\mathbf{y}}A_{0}\delta\left(  x\right)
\delta\left(  y\right)  \delta\left(  z\right)  $ is%
\begin{equation}
\mathbf{E}^{\left(  1\right)  }\left(  \rho_{0}\right)  =\frac{A_{0}k_{\rho
}^{2}R_{n}^{\prime}}{4\omega\varepsilon_{1}}e^{-p_{1}y}\left\{  \left(
\widehat{\mathbf{x}}\frac{x}{\rho_{0}}+\widehat{\mathbf{z}}\frac{z}{\rho_{0}%
}\right)  H_{0}^{\left(  2\right)  \prime}\left(  k_{\rho}\rho_{0}\right)
-\widehat{\mathbf{y}}\frac{\left(  k_{1}^{2}+p_{1}^{2}\right)  H_{0}^{\left(
2\right)  }\left(  k_{\rho}\rho_{0}\right)  }{k_{\rho}\sqrt{k_{1}^{2}-k_{\rho
}^{2}}}\right\}  ,\nonumber
\end{equation}
where $R_{n}^{\prime}=N^{E}/\left(  \partial Z^{E}/\partial k_{\rho}\right)
$, $H_{0}^{\left(  2\right)  \prime}\left(  \alpha\right)  =\partial
H_{0}^{\left(  2\right)  }\left(  \alpha\right)  /\partial\alpha$, and
$\rho_{0}=\sqrt{x^{2}+z^{2}}$. The term $e^{-p_{1}y}$ leads to exponential
decay away from the graphene surface on the proper sheet ($\operatorname{Re}%
\left(  p_{n}\right)  >0,n=1,2$). The surface wave mode may or may not lie on
the proper Riemann sheet, depending on the value of surface conductivity, as
described below. In general, only modes on the proper sheet directly result in
physical wave phenomena, although leaky modes on the improper sheet can be
used to approximate parts of the spectrum in restricted spatial regions, and
to explain certain radiation phenomena \cite{TO1963}.

\subsubsection{Transverse-Electric Surface Waves}

Noting that $p_{2}^{2}-p_{1}^{2}=k_{0}^{2}\left(  \mu_{1}^{r}\varepsilon
_{1}^{r}-\mu_{2}^{r}\varepsilon_{2}^{r}\right)  $, where $\mu_{n}^{r}$ and
$\varepsilon_{n}^{r}$ are the relative material parameters (i.e., $\mu_{n}%
=\mu_{n}^{r}\mu_{0}$ and $\varepsilon_{n}=\varepsilon_{n}^{r}\varepsilon_{0}$)
and $k_{0}^{2}=\omega^{2}\mu_{0}\varepsilon_{0} $ is the free-space
wavenumber, then if $M=1$ ($\mu_{1}^{r}=\mu_{2}^{r}=\mu_{r}$) the TE
dispersion equation (\ref{zh}) can be solved for the radial surface wave
propagation constant, yielding%
\begin{equation}
k_{\rho}=k_{0}\sqrt{\mu_{r}\varepsilon_{1}^{r}-\left(  \frac{\left(
\varepsilon_{1}^{r}-\varepsilon_{2}^{r}\right)  \mu_{r}+\sigma^{2}\eta_{0}%
^{2}\mu_{r}^{2}}{2\sigma\eta_{0}\mu_{r}}\right)  ^{2}}. \label{ksw1}%
\end{equation}
If, furthermore, $N=1$ ($\varepsilon_{1}^{r}=\varepsilon_{2}^{r}%
=\varepsilon_{r}$), then (\ref{ksw1}) reduces to
\begin{equation}
k_{\rho}=k_{0}\sqrt{\mu_{r}\varepsilon_{r}-\left(  \frac{\sigma\eta_{0}\mu
_{r}}{2}\right)  ^{2}}. \label{ksw2}%
\end{equation}
For the case $M\neq1$, then (\ref{zh}) leads to%
\begin{equation}
k_{\rho}=k_{0}\sqrt{\mu_{1}^{r}\varepsilon_{1}^{r}-\frac{1}{\left(
M^{4}-1\right)  ^{2}}\left(  M^{2}\sigma\eta_{0}\mu_{2}^{r}\mp\sqrt{\left(
\sigma\eta_{0}\mu_{2}^{r}\right)  ^{2}-\left(  M^{4}-1\right)  \left(
\varepsilon_{1}^{r}\mu_{1}^{r}-\varepsilon_{2}^{r}\mu_{2}^{r}\right)
}\right)  ^{2}}. \label{ksw3}%
\end{equation}

Considering the special case of graphene in free-space, setting $\varepsilon
_{1}^{r}=\varepsilon_{2}^{r}=\mu_{1}^{r}=\mu_{2}^{r}=1$,
\begin{equation}
k_{\rho}=k_{0}\sqrt{1-\left(  \frac{\sigma\eta_{0}}{2}\right)  ^{2}}.
\label{tesw}%
\end{equation}
If $\sigma$ is real-valued ($\sigma=\sigma^{\prime}$) and $\left(
\sigma^{\prime}\eta_{0}/2\right)  ^{2}<1$, then a fast propagating mode
exists, and if $\left(  \sigma^{\prime}\eta_{0}/2\right)  ^{2}>1$ the wave is
either purely attenuating or growing in the radial direction. However, in both
cases $p_{n}=p_{0}=\sqrt{\left(  k_{\rho}\right)  ^{2}-k_{0}^{2}}%
=-j\sigma^{\prime}\omega\mu_{0}/2$ from (\ref{zh}), and so $\operatorname{Re}%
\left(  p_{0}\right)  >0$ is violated. Therefore, for isolated graphene with
$\sigma$ real-valued (i.e., when $\sigma=\sigma_{\min}$, at low temperatures
and small $\mu_{c}$), all TE modes are on the improper Riemann sheet. The fast
leaky mode may play a role in radiation from the structure.

If the conductivity is pure imaginary, $\sigma=j\sigma^{\prime\prime}$, then
$k_{\rho}>k_{0}$ and a slow wave exists. In this case $p_{0}=\sigma
^{\prime\prime}\omega\mu_{0}/2$ and if $\sigma^{\prime\prime}>0$ then
$\operatorname{Re}\left(  p_{0}\right)  >0$ and the wave is a slow surface
wave on the proper sheet. This will occur when the interband conductivity
dominates over the intraband contribution, as described in \cite{MZPRL2007}.
However, if $\sigma^{\prime\prime}<0$, which occurs when the intraband
contribution dominates, the mode is exponentially growing in the vertical
direction and is a leaky wave on the improper sheet.

More generally, for complex conductivity,
\begin{equation}
p_{0}=\frac{-j\sigma\omega\mu_{0}}{2}=\left(  \sigma^{\prime\prime}%
-j\sigma^{\prime}\right)  \frac{\omega\mu_{0}}{2},
\end{equation}
and therefore if $\sigma^{\prime\prime}<0$ the mode is on the improper sheet,
whereas if $\sigma^{\prime\prime}>0$ a surface wave on the proper sheet is obtained.

\subsubsection{Transverse-Magnetic Surface Waves}

TM waves are governed by the dispersion equation (\ref{ze}). For general
material parameters this relation is more complicated then for the TE case,
and so here we concentrate on an isolated graphene surface ($\varepsilon
_{1}^{r}=\varepsilon_{2}^{r}=\mu_{1}^{r}=\mu_{2}^{r}=1$). Then, $p_{0}%
=-j2\omega\varepsilon_{0}/\sigma$ and
\begin{equation}
k_{\rho}=k_{0}\sqrt{1-\left(  \frac{2}{\sigma\eta_{0}}\right)  ^{2}}.
\label{tmsw}%
\end{equation}
If $\sigma$ is real-valued then $\operatorname{Re}\left(  p_{0}\right)  >0$ is
violated and TM modes are on the improper Riemann sheet.

If conductivity is pure imaginary, then $k_{\rho}>k_{0}$ and a slow wave
exists. Since $p_{0}=-2\omega\varepsilon_{0}/\sigma^{\prime\prime}$, if
$\sigma^{\prime\prime}>0$ then the wave is on the improper sheet, and if
$\sigma^{\prime\prime}<0$ the mode is a slow surface wave on the proper sheet.
For complex conductivity,
\begin{equation}
p_{0}=\frac{-j2\omega\varepsilon_{0}}{\sigma}=\frac{-2\omega\varepsilon_{0}%
}{\left\vert \sigma\right\vert ^{2}}\left(  \sigma^{\prime\prime}%
+j\sigma^{\prime}\right)  ,
\end{equation}
and therefore if $\sigma^{\prime\prime}<0$ (intraband conductivity dominates)
the mode is a surface wave on the proper sheet, whereas if $\sigma
^{\prime\prime}>0$ (interband conductivity dominates) the mode is on the
improper sheet.

In summary, for isolated graphene a proper TE surface wave exists if
$\sigma^{\prime\prime}>0$, resulting in the radial wavenumber (\ref{tesw}),
and a proper TM surface wave with wavenumber (\ref{tmsw}) is obtained for
$\sigma^{\prime\prime}<0$. The same conclusions hold for graphene in a
homogeneous dielectric. For graphene with $\sigma^{\prime\prime}=0$, no
surface-wave propagation is possible. Note that this only refers to
wave-propagation effects; dc or low-frequency transport between electrodes can
occur, leading to electronic device possibilities. In fact, in devices not
based on wave phenomena, the absence of surface waves is usually beneficial,
as spurious radiation and coupling effects, and the associated degradation of
device performance, are often associated with surface wave excitation.

The degree of confinement of the surface wave to the graphene layer can be
gauged by defining an attenuation length $\zeta$, at which point the wave
decays to $1/e$ of its value on the surface. For graphene embedded in a
homogeneous medium characterized by $\varepsilon$ and $\mu$, $\zeta
^{-1}=\operatorname{Re}\left(  p\right)  $, leading to $\zeta^{TE}%
=2/\sigma^{\prime\prime}\omega\mu$ ($\sigma^{\prime\prime}>0$) and $\zeta
^{TM}=-\left\vert \sigma\right\vert ^{2}/2\omega\varepsilon\sigma
^{\prime\prime}$ ($\sigma^{\prime\prime}<0$). When normalized to wavelength,%
\begin{align}
\frac{\zeta^{TE}}{\lambda}  &  =\frac{1}{\pi\eta\sigma^{\prime\prime}},\text{
\ \ \ \ \ \ }\sigma^{\prime\prime}>0\label{atnte}\\
\frac{\zeta^{TM}}{\lambda}  &  =-\frac{\eta\left\vert \sigma\right\vert ^{2}%
}{4\pi\sigma^{\prime\prime}},\ \ \ \ \ \sigma^{\prime\prime}<0. \label{atntm}%
\end{align}
Obviously, strong confinement arises from large imaginary conductivity, as
would be expected.

\section{Results}

In this section, some results are shown for surface wave characteristics of
graphene in the microwave and infrared regimes. In all cases $\Gamma=0.11$
meV, $T=300$ K, and an isolated graphene surface (i.e., when the surrounding
medium is vacuum) is considered. The value of the scattering rate is chosen to
be approximately the same as for electron-acoustic phonon interactions in
single-wall carbon nanotubes \cite{JDD1993}.

We first consider the case of zero chemical potential at microwave and
far-infrared wave frequencies. Fig. 2 shows the complex conductivity, TM
surface-wave wavenumber (\ref{tmsw}), and attenuation length (\ref{atntm}). In
this case the intraband conductivity is dominant over the interband
contribution, and so $\sigma^{\prime\prime}<0$, such that only a TM surface
wave can exist. The dispersion of the complex conductivity follows simply from
the Drude form (\ref{intraLR}). At low frequencies the TM surface wave is
poorly confined to the graphene surface ($\zeta^{TM}/\lambda\gg1$), and
therefore it is lightly damped and relatively fast (i.e., $k_{\rho}^{TM}%
/k_{0}\simeq1$). As frequency increases into the far-infrared, the surface
wave becomes more tightly confined to the graphene layer, but becomes slow as
energy is concentrated on the graphene surface.

The conductivity can be varied by adjusting the chemical potential, which is
governed by the carrier density via (\ref{cd}). The carrier density can be
changed by either chemical doping or by the application of a bias voltage.
Fig. 3 shows the interband and intraband conductivity as a function of
chemical potential for $\omega=6.58$ $\mu$eV ($\omega=10$ GHz). As expected,
the conductivity increases with increasing chemical potential, associated with
a higher carrier density $n_{s}$. Because intraband conductivity is dominant,
$\sigma^{\prime\prime}$ remains negative as chemical potential is varied, and
therefore only a TM surface wave may propagate. The insert shows the real part
of the intraband conductivity (the dominant term) on a linear scale, showing
the linear dependence of $\sigma$ on $\mu_{c}$.

Since the TM surface-wave is not well confined to the graphene surface at
lower microwave frequencies, it undergoes little dispersion with respect to
chemical potential. This is illustrated in Fig. 4, where the TM wavenumber and
attenuation length are shown as a function of chemical potential at
$\omega=6.58$ $\mu$eV. Because of the simple form for the TM surface-wave
wavenumber (\ref{tmsw}), the wavenumber and attenuation length merely follow
the conductivity profile.

However, at infrared frequencies moderate changes in the chemical potential
can significantly alter graphene's conductivity, and, significantly, change
the sign of its imaginary part. Fig. 5 shows the various components of the
graphene conductivity at $\omega=0.263$ eV ($\omega=400$ THz), and in Fig. 6
the total conductivity (interband plus intraband) is shown. From
(\ref{inter}), at $T=\Gamma=0$ an abrupt change in $\sigma_{inter}$ occurs
when $2\left\vert \mu_{c}\right\vert =\hslash\omega$, which in this case is
$\left\vert \mu_{c}\right\vert =0.132$ eV, denoted by the vertical dashed
lines in the figures. Since the associated Fermi temperature is several
thousand K, the $T=0$ behavior qualitatively remains the same at $300$ K,
although the discontinuity is softened due to the higher temperature. It can
be seen that for $\left\vert \mu_{c}\right\vert $ less than approximately
$0.13$ eV, $\sigma_{inter}^{\prime\prime}$ dominates over $\sigma
_{intra}^{\prime\prime}$ and $\sigma^{\prime\prime}>0$, so that only a proper
TE surface wave mode exists. Outside of this range, only a TM surface wave
propagates. This is shown in Fig. 7, where it can be seen that the TM mode is
moderately dispersive with chemical potential, especially in the region of the
sign change in $\sigma^{\prime\prime}$ near $\left\vert \mu_{c}\right\vert
\simeq0.132$ eV, since the TM surface wave is fairly well confined to the
graphene surface ($\zeta^{TM}/\lambda\leq10^{-2}$). \ In the region where
$\sigma^{\prime\prime}$ is positive, the TE mode exists but is poorly confined
to the graphene surface, and so it is essentially nondispersive and very
lightly damped. The oscillatory behavior of the attenuation length $\zeta
^{TE}$ follows simply from the form of $\sigma^{\prime\prime}$ via
(\ref{atnte}).

In Fig. 8 the conductivity is shown as a function of frequency in the infrared
regime for a fixed value of chemical potential, $\mu_{c}=0.1$ eV, at $300$ K.
The dispersion behavior of the conductivity follows simply from (\ref{intraLR}%
) and (\ref{inter}). The point $2\left\vert \mu_{c}\right\vert =\hslash\omega$
occurs at $\omega=0.2$ eV ($\omega\simeq301.6$ THz), whereupon the interband
contribution dominates and $\sigma^{\prime\prime}$ becomes positive (the
intraband contribution varies as $\omega^{-1}$ for $\omega\gg\Gamma$, and so
becomes small at sufficiently high infrared frequencies). For comparison, the
$T=0$ result is also shown.

Fig. 9 shows the TM and TE wavenumbers and attenuation lengths for the
conductivity profile given in Fig. 8. For $\omega<0.2$ eV only a TM surface
wave exists since $\sigma^{\prime\prime}<0$, and for $\omega>0.2$ eV only a TE
surface wave exists due to $\sigma^{\prime\prime}>0$. From Fig. 8 it is clear
that above $\omega=0.1$ eV, $\left\vert \sigma\right\vert \eta_{0}/2\ll1$, and
so%
\begin{align}
k_{\rho}^{TM} &  =k_{0}\sqrt{1-\left(  \frac{2}{\sigma\eta_{0}}\right)  ^{2}%
}\simeq-jk_{0}\left(  \frac{2}{\sigma\eta_{0}}\right)  ,\label{krtma}\\
k_{\rho}^{TE} &  =k_{0}\sqrt{1-\left(  \frac{\sigma\eta_{0}}{2}\right)  ^{2}%
}\simeq k_{0}\left(  1-\frac{1}{2}\left(  \frac{\sigma\eta_{0}}{2}\right)
^{2}\right)  .\label{krtea}%
\end{align}
The TM mode is tightly confined to the graphene surface, and from
(\ref{krtma}) it can be seen that in vicinity of the transition point at
$\omega=0.2$ eV, where $\sigma^{\prime\prime}\simeq0$, $k_{\rho}^{\prime
\prime}$ will be large and the TM mode will be highly damped. Just to the
right of the transition, from (\ref{krtea}) the TE wavenumber is predominately
real, and so the TE mode is very lightly damped. From (\ref{atnte}) it is
clear that $\zeta^{TE}/\lambda\gg1$, and so the TE mode is not well confined
to the surface in this frequency range.

\section{Conclusions}

An exact solution has been obtained for the electromagnetic field due to an
electric current near a surface conductivity model of graphene. Dyadic Green's
functions have been presented in terms of Sommerfeld integrals, plane-wave
reflection and transmission coefficients have been provided, and surface-wave
propagation on graphene has been discussed in the microwave and infrared
regimes. The relative importance of interband and intraband contributions for
surface wave propagation has been emphasized.

\newpage

\textbf{Figure Captions:}

\bigskip

Fig. 1. (a) Depiction of graphene (top view), where the small circles denote
carbon atoms, and (b) graphene characterized by conductance $\sigma$ at the
interface between two dielectrics (side view).

Fig. 2. Complex conductivity, TM surface-wave wavenumber, and attenuation
length for $\mu_{c}=0$ at $300$ K in the microwave through far-infrared
frequency range ($\omega=15.2$ GHz to $15.2$ THz).

Fig. 3. Interband and intraband conductivity as a function of chemical
potential at $T=300$ K, $\omega=6.58$ $\mu$eV ($\omega=10$ GHz). Note the
logarithmic scale; the insert shows the real part of the intraband
conductivity on a linear scale, showing the linear dependence of $\sigma$ on
$\mu_{c}$.

Fig. 4. Attenuation length and surface-wave wavenumber for the TM mode as a
function of chemical potential at $T=300$ K, $\omega=10$ GHz ($6.58$ $\mu$eV).
The corresponding conductivity profile is shown in Fig. 3.

Fig. 5. Interband and intraband conductivity as a function of chemical
potential at $T=300$ K, $\omega=0.263$ eV ($\omega=400$ THz). The dashed
vertical lines represent the point $2\left\vert \mu_{c}\right\vert
=\hslash\omega$.

Fig. 6. Total (interband plus intraband) conductivity as a function of
chemical potential at $T=300$ K, $\omega=0.263$ eV ($\omega=400$ THz). The
dashed vertical lines represent the point $2\left\vert \mu_{c}\right\vert
=\hslash\omega$.

Fig. 7. Attenuation length and surface-wave wavenumbers as a function of
chemical potential at $T=300$ K, $\omega=0.263$ eV ($\omega=400$ THz). The
corresponding conductivity profile is shown in Figures 5 and 6. TE and TM
modes are shown, although only portions where wavenumbers lie on the proper
Riemann sheet are provided.

Fig. 8. Total conductivity (interband plus intraband) as a function of
frequency at $T=300$ K and $\mu_{c}=0.1$ eV at infrared frequencies
($\omega=0.1$ eV$\simeq151.2$ THz). The $T=0$ result is also shown for comparison.

Fig. 9. Attenuation length and surface-wave wavenumbers as a function of
frequency at $\mu_{c}=0.1$ eV, $T=300$ K. The corresponding conductivity
profile is shown in Fig. 8.

\end{document}